\documentclass{aa} 
\usepackage{natbib}
\usepackage{graphicx}
\usepackage{txfonts}

\def\Teff{T_\mathrm{eff}}
\def\muHz{$\mu$Hz}
\def\note #1]{{\bf #1]}}

\begin{document}
   \title{A theoretical approach for the interpretation of pulsating PMS intermediate-mass stars}

   \subtitle{}

   \author{A.~Ruoppo\inst{1,2}\and
M.~Marconi\inst{1}\and
J. P.~Marques\inst{3,4,5} \and
M. J. P. F. G.~Monteiro\inst{4,5}\and
J.~Christensen-Dalsgaard\inst{6}\and
F.~Palla \inst{7}\and
V.~Ripepi \inst{1}
}

   \offprints{A.~Ruoppo}

   \institute{
INAF-Osservatorio Astronomico di Capodimonte,
Via Moiariello 16, 80131, Napoli, Italy \\
\email{ruoppo@na.astro.it, marconi@na.astro.it, ripepi@na.astro.it}
\and
Dipartimento di Scienze Fisiche, Universit\`a Federico II, Complesso
Monte S. Angelo, 80126, Napoli, Italy
\and
Grupo de Astrof\'{\i}sica da Universidade de Coimbra, Departamento de Matem\'atica - FCTUC, Portugal
\and
Departamento de Matem\'atica Aplicada, Faculdade de Ci\^encias da Universidade do Porto, Portugal 
\and
Centro de Astrof\'{\i}sica da Universidade do Porto, Rua das Estrelas, 4150-762 Porto, Portugal \\
\email{jmarques@astro.up.pt, mjm@astro.up.pt}
\and
DASC and Institut for Fysik og Astronomi, Aarhus Universitet, Denmark\\
\email{jcd@phys.au.dk}
\and
INAF-Osservatorio Astrofisico di Arcetri, Largo E. Fermi, 5, I-50125
Firenze, Italy \\
\email{palla@arcetri.astro.it}
             }

   \date{}


  \abstract
   {The investigation of the pulsation properties of pre-main-sequence intermediate-mass stars is a promising tool to evaluate the intrinsic properties of these stars and to constrain current evolutionary models. Many new candidates of this class 
have been discovered during the last decade and very  accurate data are expected from space observations obtained for example with the CoRoT satellite.}
   {In this context  we aim at developing  a theoretical approach for the interpretation of observed frequencies, both from the already available ground-based observations and from the future more accurate and extensive CoRoT results.} 
   {To this purpose we have started a project devoted to the computations of fine and extensive grids of asteroseismic models  of intermediate mass pre-main-sequence stars. The obtained frequencies are used to derive an analytical relation between the large frequency separation and the stellar luminosity and effective temperature and to develop a tool to compare theory and observations in the echelle diagram. }
   {The predictive capabilities of the proposed method are verified through the application to two test stars. As a second step, we apply the procedure to two true observations from multisite campaigns and we are able to constrain their stellar parameters, in particular the mass, in spite of the small number of frequencies.}
   {We expect that with a significantly higher number of frequencies both the stellar mass and  age could be constrained and, at the same time, the physics of the models could be tested.}
\keywords{stars: variables:  $\delta$ Scuti  -- stars:  oscillations --
	   stars: pre-main sequence --   stars: fundamental parameters --
               }

\maketitle
%
\section{Introduction}

Pre-main-sequence (PMS) $\delta$~Scuti are intermediate-mass ($ 1.5 M_{\odot}< M <4 M_{\odot}$) stars
that cross  the pulsation instability 
strip of more evolved classical pulsators during their evolution 
toward the main sequence (MS).
The first evidence of $\delta$ Scuti pulsation in PMS  stars was provided by  \citet{breger}, who discovered two candidates 
in the young cluster NGC 2264.
More than 20 years elapsed since this discovery before other authors confirmed the occurrence of this kind of pulsation
observing the Herbig Ae stars HR5999 \citep{kurtz} and HD104237 \citep{donati}. In the same years \citet{marconi} 
obtained the first theoretical instability strip for PMS intermediate-mass pulsators by computing  an extensive and detailed set of nonlinear convective models for the first three radial modes, along the PMS evolutionary tracks by \citet{palla93}.
More recently \cite{grigahcene} have produced a theoretical instabily strip for
 PMS stars for the first seven radial modes.
\cite{marconi} also pointed out that the interior
structure of PMS stars entering the instability strip differs
significantly from that of more evolved main-sequence stars (with the
same mass and temperature), even though the structure of the envelope is
similar. This property was subsequently confirmed by \citet{suran} who
made a comparative study of the seismology of a 1.8 $M_{\odot}$ PMS
and post-MS star and found that the unstable frequency range
is roughly the same for PMS and post-MS stars, but that some
non-radial modes are very sensitive to the deep internal structure of
the star. In particular, it is possible to discriminate between the
PMS and post-MS stage using differences in the oscillation frequency
distribution in the low frequency range ($g$ modes, see also
\citealt{templeton}). \par
The theoretical analysis by \citet{marconi} stimulated an increasing 
observational interest for this class of pulsators and the current number of known or suspected candidates amounts to about 30 stars.
An updated list is available at
{\small\tt http://ams.astro.univie.ac.at/pms\_corot.php}, and there are recent reviews by
\citet{zwintz}, \citet{marconi04a} and \citet{ripepi06}. 
 However, only a few stars have been studied in detail, so that the overall
properties of this class of variables are still poorly determined.
In this context our group has started a systematic monitoring program
 of Herbig Ae stars with spectral types from A to F2--3
with the following aims: 1) to identify the largest number of
pulsating objects in order to observationally determine the boundaries
of the instability strip for PMS $\delta$~Scuti pulsation; 2) to study
in detail through multisite campaigns selected objects showing
multiperiodicity (see \citealt{marconi01}; \citealt{h254,v351,ipper}; \citealt{v346}; \citealt{bernabei}) 
which are the best potential candidates for asteroseismology.\\

In order to establish a theoretical approach for the interpretation of these data, 
as well as to prepare the basis for the interpretation of future 
more accurate satellite observations expected from CoRoT \citep{baglin03}, in the context of the CoRoT/ESTA
collaboration ({\small\tt http://www.astro.up.pt/corot/}) we have started a theoretical project 
devoted to the computation of asteroseismic models of intermediate 
mass PMS stars covering a wide range of stellar masses. In this paper 
we present our first results based on an extensive grid of evolutionary 
PMS models computed with the CESAM code \citep{morel}, with updated physics \citep{marques} relevant for PMS modelling.
The Aarhus linear adiabatic non-radial pulsation code ADIPLS 
({\small\tt http://astro.phys.au.dk/$\sim$jcd/adipack.n/}) has been applied
to the models of the grid in order to build a reference base to reproduce the observed pulsation frequencies.

The paper is organized as follows:
in Section~2 we present the evolutionary PMS models, we introduce the adopted non-radial pulsation code and report  
on the predicted pulsation frequencies;
in Section~3 the computed  large separation and its dependence on luminosity and effective temperature are discussed;
in Section~4 we present a method to reproduce the observed 
frequencies and a theoretical test of the predictive capabilities of the method; 
in Section~5 we compare the theoretical results with the  observations 
from multisite campaigns of the stars V351~Ori and IP~Per. 
Finally the Conclusions close the paper.

\section{Reference grid of models with pulsation frequencies}

For exploring the pulsation properties of intermediate-mass PMS stars, 
we decided to build an extensive grid of models and frequencies to provide a reference on the seismic characteristics of PMS stars across the HR diagram.

\subsection{PMS evolution models}

The stellar models have been computed with the CESAM 
code \citep{morel} that is optimized for asteroseismology.
The precise calculation of oscillation frequencies ($p$ and $g$ modes) requires numerically precise and adequately dense meshes with the quantities describing the internal structure and entering the oscillation equations (e.g \citealt{cd}).
The physical and numerical properties of these models are discussed in detail in the paper by \cite{morel} with an updated version of the physics being used as in \cite{marques}.

%

From the CESAM PMS evolutionary tracks {\bf (see Fig.~\ref{fig4})} we selected 56
models for stellar masses varying from 1.6 to 4.0 $M_{\odot}$ with a 
step of 0.2 $M_{\odot}$.
For each selected mass, we consider  from two to four different effective 
temperatures in the range 6\,000 to 10\,000~K, along the corresponding evolutionary track.
We report in Table~\ref{tab1} the physical properties of the selected 
PMS models, which cover the empirical HR diagram location of known 
pulsating PMS intermediate-mass stars.

%
\begin{table}[h!]
 \caption{Physical properties of the selected PMS models. \label{tab1}}
\begin{tabular}{lcccrr}
\hline\hline\noalign{\smallskip}
Model & Mass & Radius & Luminosity & $\Teff$ & Age \\
      & $\rm M/M_{\odot}$ & $\rm R/R_{\odot}$ & $\rm L/L_{\odot}$
      & K & Gy \\
\hline\noalign{\smallskip}
mod1   &1.6 &1.78 &7.18   &7078  &11.21 \\
mod2   &1.6 &1.52 &6.72   &7548  &16.82 \\
mod3   &1.8 &2.32 &15.23  &7493  &7.86 \\
mod4   &1.8 &1.78 &9.70   &7648  &9.83 \\
mod5    &1.8 &2.13 &14.83  &7773  &8.30  \\
mod6   &1.8 &1.61 &11.59  &8401  &11.80 \\
mod7    &2.0 &2.75 &21.93  &7539  &5.93 \\
mod8   &2.0 &1.88 &14.27  &8184  &7.41 \\
mod9   &2.0 &1.70 &18.40  &9186  &8.89 \\
mod10   &2.2 &3.67 &18.70  &6273  &4.07 \\
mod11   &2.2 &3.10 &30.04  &7681  &4.66 \\
mod12  &2.2 &1.99 &20.74  &8749  &5.82 \\
mod13  &2.2 &1.77 &27.27  &9913  &6.98 \\
mod14  &2.4 &4.14 &24.06  &6288  &3.25 \\
mod15  &2.4 &3.48 &39.04  &7739  &3.72 \\
mod16  &2.4 &2.09 &29.46  &9304  &4.65 \\
mod17  &2.4 &2.04 &37.54  &10000 &5.11 \\
mod18  &2.6 &4.55 &33.38  &6506  &2.82 \\
mod19  &2.6 &3.92 &48.15  &7688  &3.10 \\
mod20  &2.6 &3.54 &57.56  &8459  &3.24 \\
mod21  &2.6 &3.14 &66.90  &9319  &3.38 \\
mod22  &2.6 &2.18 &43.08  &10023 &4.09 \\
mod23  &2.8 &4.81 &46.39  &6876  &2.40 \\
mod24  &2.8 &4.46 &55.53  &7468  &2.51 \\
mod25  &2.8 &4.08 &66.50  &8168  &2.63 \\
mod26  &2.8 &3.23 &90.99  &9885  &2.85 \\
mod27  &3.0 &5.58 &46.31  &6381  &1.90 \\
mod28  &3.0 &4.96 &64.92  &7366  &2.08 \\
mod29  &3.0 &4.17 &91.89  &8758  &2.26 \\
mod30  &3.0 &3.74 &107.9  &9621  &2.35 \\
mod31  &3.0 &3.32 &122.2  &10547 &2.45 \\
mod32  &3.2 &6.13 &53.10  &6301  &1.57 \\
mod33  &3.2 &5.49 &74.04  &7230  &1.73 \\
mod34  &3.2 &4.67 &105.7  &8573  &1.88 \\
mod35  &3.2 &4.22 &125.5  &9417  &1.96 \\
mod36  &3.2 &3.75 &146.2  &10375 &2.04 \\
mod37  &3.4 &6.46 &69.66  &6567  &1.37 \\
mod38  &3.4 &5.76 &94.50  &7502  &1.49 \\
mod39  &3.4 &4.96 &130.4  &8762  &1.61 \\
mod40  &3.4 &4.53 &152.5  &9539  &1.67 \\
mod41  &3.4 &4.08 &176.0  &10414 &1.73 \\
mod42  &3.6 &7.29 &67.80  &6139  &1.09 \\
mod43  &3.6 &6.46 &100.3  &7194  &1.24 \\
mod44  &3.6 &5.71 &134.7  &8234  &1.33 \\
mod45  &3.6 &4.88 &180.9  &9590  &1.43 \\
mod46  &3.6 &4.44 &207.6  &10402 &1.48 \\
mod47  &3.8 &7.84 &81.15  &6194  &0.94 \\
mod48  &3.8 &6.94 &116.3  &7203  &1.06 \\
mod49  &3.8 &6.21 &152.9  &8155  &1.14 \\
mod50  &3.8 &5.40 &202.2  &9376  &1.22 \\
mod51  &3.8 &4.97 &231.2  &10106 &1.26 \\
mod52  &4.0 &8.25 &101.0  &6375  &0.83 \\
mod53  &4.0 &7.30 &140.7  &7369  &0.92 \\
mod54  &4.0 &6.58 &180.5  &8253  &0.98 \\
mod55  &4.0 &5.80 &232.9  &9366  &1.05 \\
mod56  &4.0 &5.40 &263.8  &10023 &1.08 \\
\noalign{\smallskip}
\hline\hline
\end{tabular}
\end{table}

\subsection{Calculation of the oscillation frequencies}

In order to explore the pulsation frequencies that characterize the selected  
PMS models, associated to both 
radial and non-radial modes, we applied the ADIPLS code.
The pulsation frequencies are obtained as solutions of a set
of linear equations for small perturbations {\bf (for details see \citet{unno}).}
{\bf The physical and numerical assumptions used for the computation are 
discussed in detail by} \cite{cd82,cd,cd94}.\\

\begin{figure}
   \centering
   \includegraphics[width=8cm]{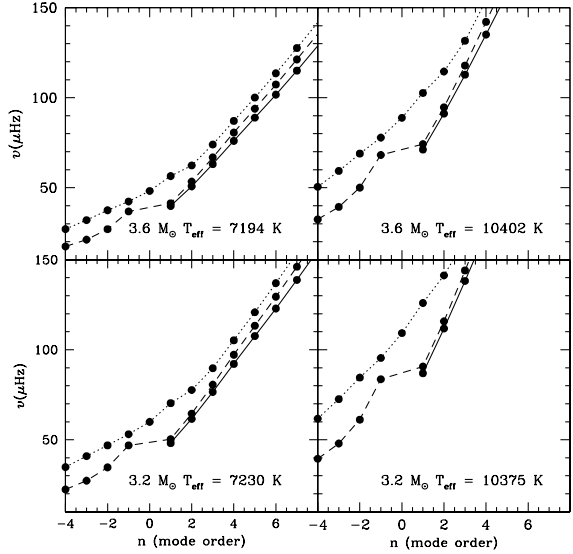}
      \caption{Predicted frequencies as a function of the radial order $n$ for models at
      3.2~$M_{\odot}$ (bottom panel: {\em mod33} left, {\em mod36} right), 
      and 3.6~$M_{\odot}$ (upper panel: {\em mod43} left, {\em mod46} right). 
      Symbols are connected by a full line in the
      case of $l$=0, by a dashed line for $l$=1 and by a dotted 
      line for $l$=2.
      The effective temperatures of the models are indicated for each mass.
      \label{fig2}}   
   \end{figure}

In this work we concentrate mainly on $p$ modes, i.e. acoustic modes,
with a spherical harmonic degree $l=0,1,2$ and radial order $n=0,1,...,20$. 
We only consider very low degree modes because these are thought to be more likely detectable
(e.g. \citealt{dz77,CD82}).
From a first inspection of the results we note that the predicted range of $p$-mode frequencies moves toward higher values as the stellar mass decreases, 
at fixed effective temperature, and as the effective temperature increases, 
at fixed stellar mass (see Fig.~\ref{fig2}). This behaviour confirms the well 
known relation between the pulsation frequency and the stellar mean density.

%
\section{The large frequency separation} \label{sec:lfs}

The pattern of regular separations between frequencies of modes of 
different degree and order is a characteristic feature expected in $p$-mode stellar oscillations.
The frequency separations depend on the internal structure of the star 
so by measuring them we gain valuable diagnostics on the stellar interior.

   \begin{figure}
   \centering
   \includegraphics[width=8cm]{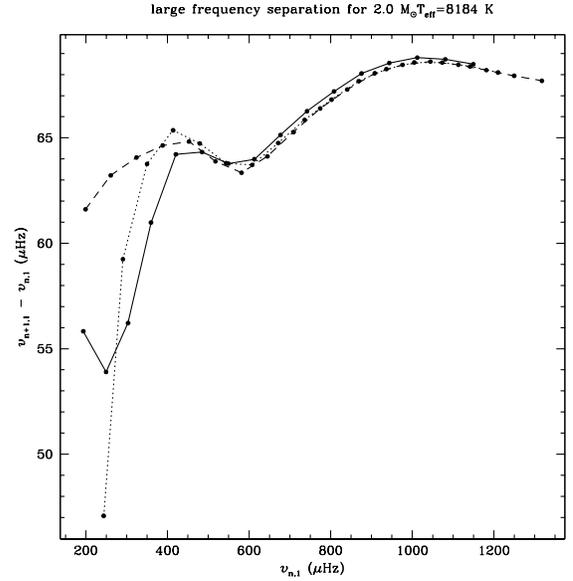}
      \caption{Large frequency separation as a function of frequency for the 
      model with $M=2.0 M_\odot$ and $\Teff=8184$~K, for $1 < n \le 18$. 
      Symbols are connected by a full line in the
      case of $l$=0, by a dashed line for $l$=1 and by a dotted
      line for $l$=2.
      \label{fig3}}
   \end{figure}

{\bf In particular, in the asymptotic regime}, it is convenient to use the large frequency separation defined as
\begin{equation}
\label{eq:eq2}
\Delta \nu_{n,l} \equiv \nu_{n,l} - \nu_{n{-}1,l} \simeq \Delta\nu \;,
\end{equation}
between modes of the same order{\bf , where  
$\Delta\nu \simeq (\int_0^R {{\rm d} r / c})^{-1}$ $\propto \left( {M / R^3} \right)^{1/2} $, with $c$ the sound speed in the star. }

In Fig.~\ref{fig3} we show for model {\em mod8} the behaviour of the large separation as defined in Eq.~(\ref{eq:eq2}), for $l=0, 1,2$, versus the frequency. 

{\bf On this basis} a relation
can be found between the large frequency separation,
the luminosity and the effective temperature
(the stellar mass is related to the luminosity for each effective temperature).
For this purpose, we have computed a large separation $\Delta\nu$ in the asymptotic region as the mean of all values for $5<n\le 20$.

\begin{figure}
   \centering
   \includegraphics[width=8cm]{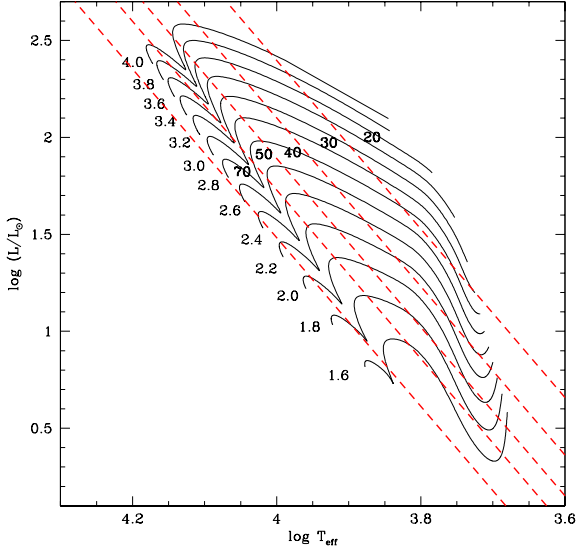}
      \caption{HR diagram showing PMS evolutionary tracks 
      computed with the CESAM code
      for the labelled stellar masses (in $M_\odot$)
      The linear fit given in eq.~(\ref{eq:eq4}) is
      represented by dashed lines, defined for 
      a constant value of $\Delta\nu$ (as labelled, in \muHz),
      from 20~\muHz\ to 70~\muHz.
      \label{fig4}}
   \end{figure}

A linear fit for all models in our grid gives the following relation:
\begin{eqnarray}
\label{eq:eq4}
\log L &=& \big(-12.81 \pm 0.02\big)
   + \big(4.35 \pm 0.05\big) \; \log \Teff - \nonumber\\
&& - \big(1.681 \pm 0.015\big) \;\log \Delta\nu \;,
\end{eqnarray}
plotted in the HR diagram in  Fig.~\ref{fig4}, 
for selected values 
of $\Delta\nu$, ranging from 20~\muHz\ to 70~\muHz.
This relation is useful to obtain an expected range of the large separation of an observed pulsating star once its position in the HR diagram
has been estimated.
In the next section we use this approach as a first step to characterise a PMS $\delta$~Scuti pulsator with the adopted non-radial pulsation models  
and the observed frequencies.

%

%
\section{How to reproduce the observed frequencies}

In this section we present a methodology to compare the observed pulsation 
frequencies with theoretical ones.
This comparison allows us to obtain information 
on the stellar structure and physical parameters, such as the mass, the radius, the effective temperature and the age.
For simulating typical available data for observed stars, we ``theoretically built'' two PMS test stars.
The first one, {\em Star1}, corresponds to a model computed using the 
STAROX code \citep{roxburgh} as determined for Task~1 of CoRoT/ESTA, while the second one, {\em Star2}, was obtained 
with the CESAM code \citep{morel} after integration of the birthline from \cite{palla93} as the initial configuration for the evolution calculated by CESAM \citep{marques06}.
The parameters characterising these models are given in Table~\ref{tab2}.

We adopted two different evolutionary codes for producing the two 
test objects (also changing the initial condition in one of these) because we need to ensure that the use of the same code for building the test cases and the grid of models used here to estimate the stellar parameters is not the dominant factor determining the success of our inversion procedure.
The physical properties of these two test stars are presented in  
Table~\ref{tab2}, whereas the corresponding theoretical frequencies, 
for both radial and non-radial $p$ and $g$ pulsation modes, 
were computed using the POSC code \citep{monteiro96} -- again a different code from the one used to calculate the frequencies of the reference grid.
The values of the frequencies, the  spherical harmonic degree $l$ and the radial order $n$, are reported in Table~\ref{tab3}.

%
\begin{table}[h!]
 \caption{Physical properties (exact values) of the two test stars. These values were only known to one member of the team.
 The quantities provided to the other team members are listed as `Observed' values. \label{tab2}}
\begin{tabular}{lccccr}
\hline\hline\noalign{\smallskip}
 &Mass &  Radius & Luminosity & $\log(\Teff)$ & Age\\
  & $M/M_{\odot}$   & $R/R_{\odot}$  & $\log(L/L_\odot)$ & K & My \\
\hline\noalign{\smallskip}
\multispan{2}{Exact values:\hfill } \\[4pt]
{\em Star1} &2.0 &1.862 & 1.1942 & 3.9251 & 0.00\\
{\em Star2} &2.4 &3.482 & 1.5915 & 3.8887 & 3.72\\
\hline\noalign{\smallskip}
\multispan{2}{`Observed' values:\hfill } \\[4pt]
{\em Star1} &  &  & 1.19$\pm$0.82 & 3.92$\pm$0.06 & \\
{\em Star2} &  &  & 1.58$\pm$0.07 & 3.88$\pm$0.05 & \\
\noalign{\smallskip}
\hline\hline
\end{tabular}
\end{table}

%
\begin{table}[h!]
 \caption{Frequencies, spherical harmonic degree, and the radial order 
as computed using the POSC code for the two test stars.
In the last column the `observed' frequencies provided as observables for the two test stars are also given. \label{tab3}}
\begin{tabular}{rcccccrr}
\hline\hline\noalign{\smallskip}
 & $\nu_{n,l}$ & $l$ & $n$ & \qquad & `Observed' $f_i$\\
 & \muHz & & & & \muHz \\
\hline\noalign{\smallskip}
{\em Star1} \\[4pt]
$ f_1$&   428.4412& 0   &    5  & & 429 $\pm$ 4.5 \\            
$ f_2$&   493.4204& 0   &    6  & & 493 $\pm$ 4.5 \\             
$ f_3$&   526.5400& 1   &    6  & & 521 $\pm$ 4.5 \\ 
$ f_4$&   553.4227& 2   &    6  & & 553 $\pm$ 4.5 \\  
$ f_5$&   591.5814& 1   &    7  & & 594 $\pm$ 4.5 \\ 
$ f_6$&   623.5623& 0   &    8  & & 623 $\pm$ 4.5 \\                 
$ f_7$&   655.9865& 1   &    8  & & 658 $\pm$ 4.5 \\  
$ f_8$&   683.0216& 2   &    8  & & 682 $\pm$ 4.5 \\    
$ f_9$&   748.6396& 2   &    9  & & 749 $\pm$ 4.5 \\ 
$ f_{10}$&754.4297& 0   &   10  & & 755 $\pm$ 4.5 \\[4pt]            
\hline\noalign{\smallskip}
{\em Star2} \\[4pt]
$ f_1$&    67.1478 &1&-2 & & 68 $\pm$ 3.5 \\
$ f_2$&    83.9724 &0& 1 & & 84 $\pm$ 3.5 \\
$ f_3$&   108.1624 &0& 2 & &108 $\pm$ 3.5 \\
$ f_4$&   112.4021 &1& 1 & &111 $\pm$ 3.5 \\
$ f_5$&   134.5490 &0& 3 & &134 $\pm$ 3.5 \\ 
$ f_6$&   168.5343 &1& 3 & &166 $\pm$ 3.5 \\
$ f_7$&   183.7885 &2& 4 & &183 $\pm$ 3.5 \\ 
$ f_8$&   188.1748 &0& 5 & &188 $\pm$ 3.5 \\
$ f_9$&196.8293 &1& 5 & &196 $\pm$ 3.5 \\ 
$ f_{10}$&224.3726 &1& 6 & &224 $\pm$ 3.5 \\ 
$ f_{11}$&240.7379 &0& 7 & &242 $\pm$ 3.5 \\ 
$ f_{12}$&265.4323 &0& 8 & &266 $\pm$ 3.5 \\
$ f_{13}$&268.3870 &2& 7 & &268 $\pm$ 3.5 \\
$ f_{14}$&294.3213 &2& 8 & &293 $\pm$ 3.5 \\ 
$ f_{15}$&296.9251 &0& 9 & &297 $\pm$ 3.5 \\ 
\noalign{\smallskip}
\hline \hline
\end{tabular}
\end{table}

We note that most of these frequencies have been verified to correspond
to modes that are actually unstable, 
through the use of nonadiabatic computations.
We also notice that only one author knew the
real stellar parameters and the mode parameters of the frequencies,
while the others tried to find the best solution reproducing the 
observed stellar parameters and frequencies, just as if they were true observed data. 
In particular, the author that knew the properties of the two test stars 
has simulated real data by shifting the frequencies, the luminosity and the effective temperature relatively to the exact value and quoting an error not always compatible with the original (exact) value. 
The frequencies for $g$ modes have also been added in one case.
All values, exact and simulated observations, are 
reported in Tables~\ref{tab2}--\ref{tab3}.

In both test cases the number of frequencies is high (10 or more) and with a few of those already in the asymptotic regime.
Present data on PMS $\delta$~Scuti stars are not so rich (only a few frequencies) but forthcoming observations from CoRoT are expected to provided several tens of frequencies per star.
To illustrate the present situation we consider later an application to two  real cases observed in  multisite campaigns.

Based on the `observed' data we have tried to establish if we could determine the basic stellar parameters.
In doing so an approach has been developed in order to best use the available data when there is no mode identification and poorly constrained stellar parameters (as is often the case for single PMS stars).
To summarize our approach the steps we follow are:
\begin{itemize}
\item combine the grid of models with the known luminosity and effective temperature to determine a range in mass and expected large frequency separation;
\item estimate the large frequency separation from the frequency data;
\item reduce the mass range by using the observed large frequency separation;
\item use a few models within the possible range of mass and large separation to reproduce the frequencies in the echelle diagram;
\item based on the best representation in the echelle diagram able to reproduce the observations we provide the preferred stellar parameters and a tentative mode identification.
\end{itemize}
This sequence is motivated by the need to use sequentially the most robust information contained in the frequencies.
This is done by reducing as much as possible the effects on the analysis of the unknown physics that is not included in the reference grid of models and frequencies.

The proposed sequence of steps is executed for our two test stars as described in the next sections.

\subsection{The HR diagram}

\begin{figure}
   \centering
   \includegraphics[width=8cm]{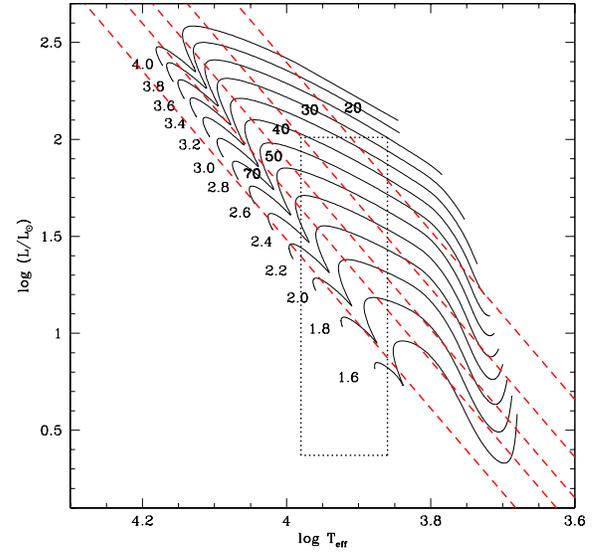}
      \caption{HR diagram position of {\em Star1} (dotted square)
      together with CESAM PMS evolutionary tracks (full lines, labelled with
      masses in $M_\odot$)
      and the linear fit (dashed lines, labelled with $\Delta \nu$ in \muHz) 
      of constant 
      large frequency separation obtained in Section~\ref{sec:lfs}.
      \label{hr1}}
   \end{figure}

\begin{figure}
   \centering
   \includegraphics[width=8cm]{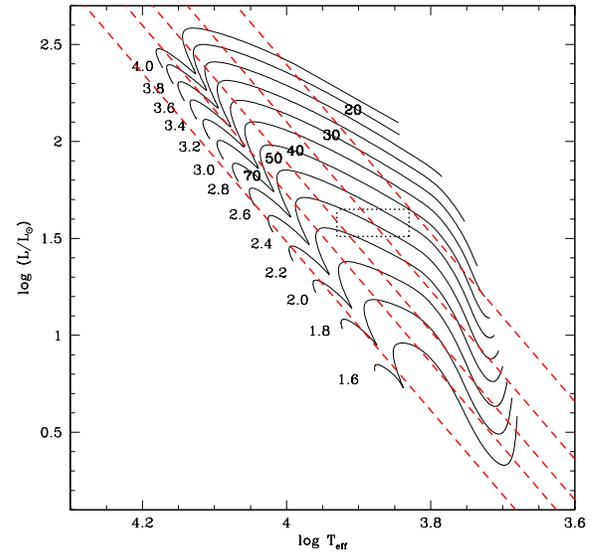}
      \caption{The same as Fig.~\ref{hr1}, but for {\em Star2}.
            \label{hr2}} 
   \end{figure}

On the basis of the reported range of luminosity and effective temperature
we show in the Figs~\ref{hr1}--\ref{hr2} the predicted 
HR diagram position of {\em Star1} and {\em Star2} respectively.
In the same plots we also display the set of PMS evolutionary tracks from CESAM for stellar masses from $1.6$ to $4.0$ solar masses, together with 
the linear fit obtained in the previous section for constant large 
separations (ranging from 20 to 70~\muHz).
These plots allow us to define the range in mass consistent with each star and 
determine the possible range of values for the large separation.
 The obtained mass and large separation ranges are:
 \begin{itemize}
\item []{\it Star1}: $1.6$ $\leq$  $M/M_{\odot}$ $\leq 3.4$ and 20~\muHz $\leq$ $\Delta\nu$ $\leq$ 80~\muHz, \\
\item []{\em Star2}: 2.2 $\leq$ $M/M_{\odot}$ $\leq$ 2.8 and 20~\muHz $\leq$ $\Delta\nu$ $\leq$ 50~\muHz.
\end{itemize}
These are yet large ranges of possible values which correspond to the uncertainty on stellar mass when only the classical observables (luminosity and effective temperature) are available.

\subsection{The large frequency separation}

In a real case the large separation would be ideally extracted directly from the power spectrum of the light curve based on the expected range reported above.
The development of an extraction method for this quantity from the power spectrum, based on the methodology developed for solar-type stars but adapted to $\delta$~Scuti PMS stars, will be described elsewhere.

For the test cases we have here, there is not a power spectrum
so an alternative analysis has to be used in order to estimate the large separation from the actual frequencies.
Based on the regularity of the spacing between
frequencies we obtain an indication of the large separations as being about
$\Delta\nu_1 \sim 65$~\muHz\
and $\Delta\nu_2\sim27$~\muHz\ for {\em Star1} and {\em Star2}, respectively. 
This approach does not provide an uncertainty but in a real case this `error' could be as high as $\sigma(\Delta\nu)\sim 5$~\muHz.

\subsection{The echelle diagram}

As it is well known, the echelle diagram is convenient to illustrate in detail the properties of the frequency spectrum. 
In this diagram the frequencies are reduced modulo 
$\Delta \nu$ by expressing them as
\begin{equation}
\nu_{n,l}= \nu_0 + k \; \Delta\nu + \tilde\nu \;,
\end{equation}
where $\Delta\nu$ is 
the large separation, $\nu_0$ is a suitably chosen reference,
and $k$ is an integer such that $\tilde{\nu}$ is between 0 and $\Delta\nu$.
The diagram is produced by plotting  $\tilde{\nu}$ on the abscissa and 
$\nu_0 + k \, \Delta\nu$ on the ordinate. 
If the asymptotic relation 
between frequencies  and large separation were 
precisely satisfied, we would obtain data points  arranged on a set of 
vertical lines corresponding to the different values of $l$.



\begin{figure}
   \centering
   \includegraphics[width=8cm]{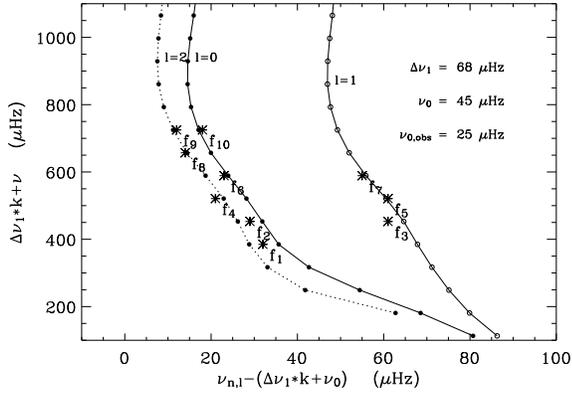}
      \caption{Echelle diagram for {\em Star1}, using the theoretical 
      frequencies computed with the Aarhus code.
      The parameters used to produce this plot are shown in the plot.
      The observational uncertainties would correspond approximately
      to the size of the symbols representing the observed
      frequencies $f_i$.\label{echelle1}}
   \end{figure}

\begin{figure}
   \centering
   \includegraphics[width=8cm]{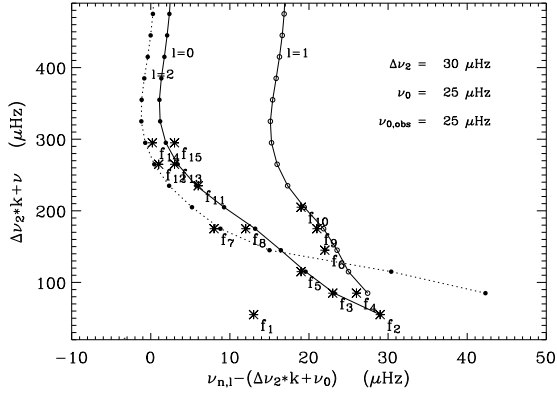}
      \caption{The same as Fig.~\ref{echelle1}, but for {\em Star2}.
 \label{echelle2}} 
   \end{figure}

 By using the best values for the large separation, determined before, 
we select all the models 
for which the intersection of the corresponding constant $\Delta\nu$  
lines with the 
evolutionary tracks are inside the assumed uncertainty 
box (see Figs~\ref{hr1}--\ref{hr2}).
For the models located at these intersections we 
use the computed frequencies to construct theoretical echelle diagrams
and search for the one that best reproduces 
(i.e. matches as many as possible of) the observed frequencies.
As a result we obtain
$M=2.0~M_{\odot}$, $\Teff=8184$~K for {\em Star1}
and $M= 2.4~M_{\odot}$, $\Teff=7739$~K for {\em Star2}.
The corresponding echelle diagrams are shown in 
Figs~\ref{echelle1}--\ref{echelle2}.  
 
\subsection{Results}

As shown in Fig.~\ref{echelle1}, for {\em Star1} all the
frequencies are reproduced with our best fit model, and we are 
able to obtain the correct spherical harmonic degree $l$ and  radial 
order $n$, for each frequency.
Our best fit model has also the right mass and an
effective temperature that differs from the true one by only 200~K.
We remind that this is the case with a poorly determined luminosity, but whose impact on the mass estimation is weak.
Based on the mode identification provided by this approach, a more detailed study of the global characteristics and the internal structure of the star could now be attempted.

Also for {\em Star2} we are able to reproduce the right mass and
effective temperature and all the 16 frequencies with the right $l$ and 
$n$ degrees (see Fig.~\ref{echelle2}).
The only exception is $f_1$ that corresponds to a $g$ mode and this is the reason why  the echelle diagram has been unable -- correctly -- to adjust this particular value, even if our  best-fitting model has a $g$ mode that fits $f_1$ within $0.005$ $\mu Hz$.

These are both ideal test cases, but they demonstrate that if enough frequencies are known with sufficiently high precision, the oscillation spectra can be interpreted using as reference a detailed grid of PMS models and their frequencies as proposed in this work.
The key assumption is that the measured values of the frequencies do not deviate strongly from the predicted values and relative spacing.
Any effect that changes significantly the linear adiabatic radial and non-radial frequencies used here as the reference may invalidate this assumption.
These can include fast rotation and strong magnetic effects on the frequencies.
Before a definite characterisation of the observed modes is reached, these effects need to be taken into account on a case-by-case basis.

The heavy-element abundance $Z$ (or metallicity) is also an important parameter of the grid (here $Z=0.02$ has been used) that has an impact on the outcome of the fitting.
This parameter can, however, be included in the proposed procedure either after the last step by re-building the echelle diagram with a model with an adequate $Z$ or by using a reference grid of models calculated with representative metallicities.

%
\section{Preliminary comparison with observations}

Having tested our theoretical approach on artificial pulsators, we may now try 
to evaluate how it can be applied to real observations.
The difficulty with ground-based observations of PMS $\delta$~Scuti stars is that these do not provide yet a large set of frequencies for each star.
But space observations have already been reported where several tens of frequencies with very high precision \citep{most}
are measured for a single star.
These will be the ideal cases for applying the approach described above.

%
\subsection{The case of V351 Ori}

As a preliminary application we considered the known multiperiodic PMS $\delta$~Scuti star V351~Ori (see \citealt{v351}).
The observed frequencies for this object, with the associated 
uncertainties, are reported in Table~\ref{v351ori}.
As noted by \citet{v351} the first four frequencies are well established, 
whereas $f_5$ is slightly less reliable.
Moreover $f_4$ and $f_5$ are quite close to $f_1$ and $f_3$ respectively.
As discussed by \citet{breger02} close frequency pairs should be investigated accurately in order to establish whether they are real or not. In particular the quoted authors suggest a method to discriminate which hypothesis (close pair or single mode with amplitude variation) is correct. Unfortunately, current data on V351 Ori are insufficient to follow this approach.
Consequently we will mainly focus on reproducing frequencies $f_1$, $f_2$ and $f_3$.

%
\begin{table}[h!]
 \caption{Observed frequencies and associated uncertainties for V351 Ori, as reported by \citet{v351}. \label{v351ori}}
\begin{tabular}{lcr}
\hline\hline\noalign{\smallskip}
 & Frequency & error\\ 
  & \muHz & \muHz \\
\hline\noalign{\smallskip}
$f_1$&181.6&0.4\\
$f_2$&165.9&0.4\\
$f_3$&147.6&0.4\\
$f_4$&183.8&0.4\\
$f_5$&148.3&0.4\\
\noalign{\smallskip}
\hline \hline
\end{tabular}
\end{table}

\begin{figure}
   \centering
   \includegraphics[width=8cm]{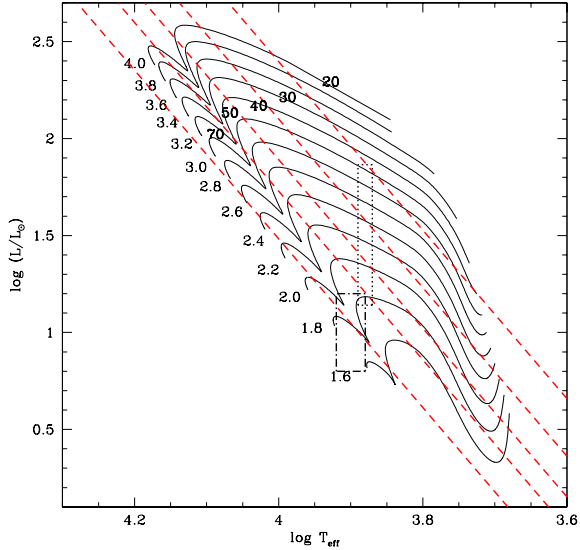}
      \caption{The same as Fig.~\ref{hr1}, but for stars V351~Ori (box with dotted square line), IP~Per (box with dot-dashed line).
\label{hrv351ori}} 
   \end{figure}

The ranges of luminosity and effective temperature as reported in the literature for V351~Ori
are plotted in the HR diagram (dotted box) in Fig.~\ref{hrv351ori} with the PMS evolution tracks. Similarly to the plots presented for the two test stars, we also report in the same diagram the lines at constant large frequency separation.
This comparison allows us to obtain a range in mass and large separation
consistent with the estimated HR diagram position of the star, namely
$1.8 < M/M_{\odot} < 3.0$ and 20~\muHz~$<\Delta\nu_a<50$~\muHz.

The next step is to identify from the frequency data if a stronger constraint can be posed on the large frequency separation.
The large separation of V351~Ori is difficult to estimate directly from the power spectrum.
Instead, we use the separations between observed frequencies to evaluate for what conditions the modes with larger amplitude (expected to be unstable low-order radial modes) are consistent with the range of $\Delta\nu$ found from the HR diagram.
By using the reference grid we can extrapolate to the asymptotic regime to find that the frequencies are consistent with a $\Delta\nu_a \sim 40$~\muHz.

\begin{figure}
   \centering
   \includegraphics[width=8cm]{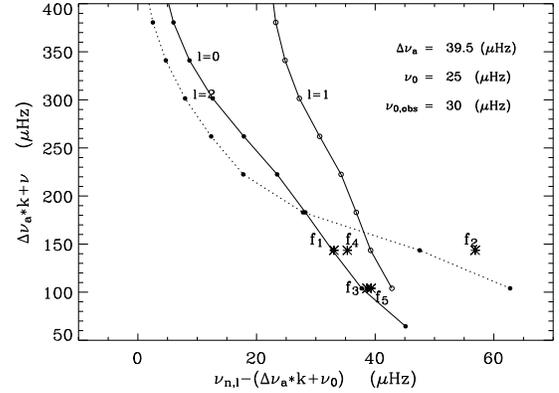}
      \caption{The same as figure \ref{echelle1}, for V351 Ori.
\label{echellev351}} 
   \end{figure}

Consequently, to find the best model that reproduces the observed 
frequencies we built the echelle diagram by varying the mass 
and the large separation within the constraints indicated above.
In doing so we assume that the mode with the largest amplitude ($f_1$) is a radial mode.
As a result, we find that the model in the grid consistent with
the observed frequencies has a mass $M=2.0~M_{\odot}$
and an effective temperature of $\Teff=7539$~K.
As shown in the echelle diagram (see Fig.~\ref{echellev351}), 
the frequencies $f_1$ and $f_3$ are associated with radial modes ($l=0$),  namely  to the first and second overtone respectively, while $f_2$ is a nonradial mode (rather close to the $l=2$ theoretical sequence).
As for $f_4$ and $f_5$ their position in the echelle diagram is between the $l=0$ and the $l=1$ sequences, but as discussed above and in particular for $f_5$, the true nature should be investigated carefully on the basis of more extensive data.\\

There are key assumptions that were made that require further verification.
One is the assumption that the highest amplitude mode ($f_1$) is a radial mode.
This is required to properly adjust the echelle diagram to the observed spectrum.
This hypothesis, or an equivalent calibration for a radial mode, requires verification from spectroscopic mode identification.

Rotation also introduces an additional difficulty as it may change
the mode identification we obtain for non-radial modes and shift the
frequencies.
In particular, for our best-fitting model the rotational splitting
should be of
the order of $8\,m/\sin{i}$~\muHz\ for an estimated rotation velocity
$v\sin{i} \simeq 1 00$~km/sec (see \citealt{balona}).
Such a relatively large splitting removes the
possibility that the separation between $f_4$ and $f_1$ and between
$f_3$
and $f_5$ is due to rotation, once the hypothesis of
radial modes for $f_1$ and $f_3$ is abandoned.
However, $f_2$ is shifted by about 10~\muHz\ from the predicted
frequency, which
may correspond to a $m\ne0$ mode, pending spectroscopic confirmation.
It should also be noted that for such a relatively rapid rotation
higher-order effects must be taken into account
\citep[e.g.,][]{gt90, dg92, sgd98}.
These lead to non-uniform splitting and shifts of modes with $m = 0$,
including the radial modes.
It is likely that such rotational frequency perturbations will be a
substantial complication in the analysis of oscillations of PMS
stars,
which typically are rapid rotators.


%
\subsection{The case of IP Per}

Another case that we may also consider as a possible preliminary application is IP~Per, which has been the target of a recent multisite campaign \citep{ipper}.
The observed frequencies are listed in Table~\ref{tab:ipper} while in Fig.~\ref{hrv351ori} the HR diagram includes the location of the star (box with dot-dashed line) and the reference grid of evolution tracks.
From this plot we can restrict the mass and large separation expected for this star to the ranges $1.6<M/M_\odot<2$ and 50~\muHz~$<\Delta\nu_b<75$~\muHz.

%
\begin{table}[h!]
 \caption{Observed frequencies and associated uncertainties for IP~Per, as reported by \citet{ipper}. \label{tab:ipper}}
\begin{tabular}{lcr}
\hline\hline\noalign{\smallskip}
 & Frequency & error\\ 
  & \muHz & \muHz \\
\hline\noalign{\smallskip}
$f_1$&264.9&1.3\\
$f_2$&400.5&1.3\\
$f_3$&352.4&1.3\\
$f_4$&558.2&1.3\\
$f_5$&333.2&1.3\\
$f_6$&277.6&1.3\\
$f_7$&107.6&1.3\\
$f_8$&487.4&1.3\\
$f_9$&602.3&1.3\\
\noalign{\smallskip}
\hline \hline
\end{tabular}
\end{table}

\begin{figure}
   \centering
   \includegraphics[width=8cm]{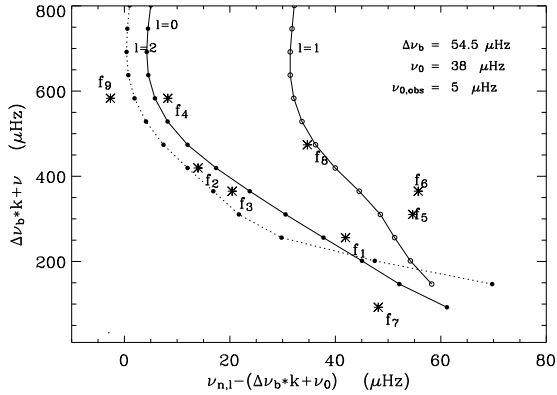}
      \caption{The same as figure \ref{echelle1}, for IP Per. 
\label{echelleipper}} 
   \end{figure}

Due to the higher number of modes, and considering that five of these have a large amplitude, the separation between observed frequencies indicates that the data are consistent, within the expected range, with $\Delta\nu_b\sim 50$~\muHz, when extrapolated to the asymptotic regime.
Following these constraints we have then used the echelle diagram to find the best stellar mass and age that reproduce as many as possible of the observed frequencies.
As a result we find a best fit model with $M=1.8~M_{\odot}$, and $\Teff=7773$~K. The corresponding echelle diagram is shown in Fig.~\ref{echelleipper}.

According to this plot, frequencies from $f_1$ to $f_4$ seem to align along the sequence for $l=0$ modes, whereas $f_5$, $f_6$ and $f_8$ are better in agreement with $l=1$ theoretical predictions, and $f_9$ could be an $l=2$ mode.
The discrepancy between the theoretical and observed frequencies is probably due to a residual uncertainty on stellar mass, intrinsic to the adopted model grid. A slightly lower mass would probably allow us to better reproduce the observed frequencies in the echelle diagram.  
In any case the frequency $f_7$ is not consistent with a $p$ mode as it lies well below the expected frequencies for the range of stellar masses of IP~Per.
This frequency could correspond to a $g$ mode, but we have not explored this possibility further here.  

\section{Conclusions}

In this work we establish a procedure based on an extended grid of models and oscillation frequencies to analyse the properties of PMS $\delta$~Scuti stars.
The procedure uses a grid of PMS evolution models to identify in the HR diagram the range of stellar mass.
From the frequencies we estimate the large frequency separation which is then used to reduce the uncertainty on stellar mass.
Finally, from the detailed analysis of the echelle diagram for the few possible combinations of mass/age, a fit of as many observed frequencies as possible is obtained.
This fit provides an initial guess on the mode parameters and a precise estimation of the stellar mass.
The underlying principle of this approach is to extract sequentially the information from the frequencies that are more robust and directly connected to the global parameters, reducing in this way as much as possible the effect on the inferences of the unknown physics in this type of stars.
We also notice that the proposed method is quite general, as it can be applied by using models computed with any evolutionary code.

In order to evaluate the validity of the proposed approach we have considered two test cases.
In both cases the result was positive, and the parameters were recovered with considerable precision even when the luminosity of the stars is poorly known.
The consistency between the true and the inferred stellar parameters for both star 1 (computed with STAROX) and star 2 (computed with CESAM), seems to indicate that the method does not depend on the adopted evolutionary code. A comparison with other evolutionary codes (e.g. FRANEC, ATON, Siess et al. 2000) will be performed when a large number of accurate frequencies, as expected from CoRoT observations, will be available for PMS $\delta$ Scuti stars.

The method we propose is mainly aimed at studying stars whose pulsation data include several tens of frequencies, as it is expected to be obtained by CoRoT in the near future.
However, we have also considered the possibility of applying it to stars whose frequency spectrum includes a smaller number of measured frequencies.
To do so we have reported the preliminary applications to two stars that have been targets of multisite campaigns: V351~Ori and IP~Per.
The uniqueness of the solution is strongly dependent on the number of frequencies being used when these are less than about ten.
However, the results illustrate the capability of this approach to constrain the stellar parameters, and in particular the stellar mass, even when a small number of frequencies is available.

For a higher number of frequencies the stellar mass and age can be constrained and the frequencies used to test the physics of the models. 
The application to a much larger number of 
frequencies for PMS pulsators (in particular as expected 
from space observations) and the adoption of finer model grids could 
considerably improve our knowledge of the stellar properties and structure in this very important phase of stellar evolution.  A comparison with other evolutionary codes (e.g. FRANEC, ATON, Siess et al. 2000), as planned in the context of the CoRoT/ESTA collaboration, is also desirable.

\begin{acknowledgements}
We thank our referee W. Weiss for his valuable comments and suggestions. 
MJM and JPM were supported in part by FCT through project {\scriptsize POCI/CTE-AST/57610/2004} from POCI, with funds from the European programme FEDER.

\end{acknowledgements}

%

\end{document}